\documentclass[aps,pre,twocolumn,showpacs,floatfix,a4paper]{revtex4}

\usepackage{graphicx}
\usepackage{amsmath}
\usepackage{amssymb}
\usepackage{bm}

\renewcommand{\vec}[1]{{\mathbf{#1}}}

\newcommand{\matr}[1]{{\mathsf{#1}}}
\newcommand{\unitmatr}{{\mathbb I}}

\newcommand{\mr}[1]{{\mathrm{#1}}}
\newcommand{\bfig}{\begin{figure}}
\newcommand{\efig}{\end{figure}}

\newcommand{\Tc}{\ensuremath{T_\mathrm{c}}}

\newcommand{\Tg}{\ensuremath{T_\mathrm{g}}}

\newcommand{\lstat}{\ensuremath{\sigma}}
\newcommand{\rbond}{\ensuremath{r_\text{b}}}

\newcommand{\rsfig}[2][0.4]{\begin{center}
                       \includegraphics*[width=#1\textwidth]{#2}
                       \end{center}
                       }

\newcommand{\figref}[1]{Fig.~\ref{#1}}

\begin{document}


\title{Static Properties of a Simulated Supercooled Polymer Melt: Structure Factors, Monomer Distributions Relative to the Center of Mass, and Triple Correlation Functions}

\author{Martin Aichele}
\affiliation{Institut f\"ur
  Physik, Johannes Gutenberg-Universit\"at, 55099 Mainz, Germany}
\affiliation{Institut Charles Sadron, 6 rue
  Boussingault, 67083 Strasbourg, France}
\author{Song-Ho Chong}
\affiliation{Laboratoire de Physique Math\'ematique et Th\'eorique,
  Universit\'e Montpellier II, 34095 Montpellier, France}
\author{J\"org Baschnagel}
\email[Corresponding author. Email: ]
  {baschnag@ics.u-strasbg.fr}
\affiliation{Institut Charles Sadron, 6 rue
  Boussingault, 67083 Strasbourg, France}
\author{Matthias Fuchs}
\affiliation{Fachbereich Physik, Universit\"at Konstanz, 
  78457 Konstanz, Germany}

\begin{abstract}
We analyze structural and conformational properties in a simulated bead-spring model of a non-entangled, supercooled polymer melt.  We explore the statics of the model via various structure factors, involving not only the monomers, but also the center of mass (CM).  We find that the conformation of the chains and the CM-CM structure factor, which is well described by a recently proposed approximation [Krakoviack {\em et al.}, Europhys.\ Lett.\ {\bf 58}, 53 (2002)], remain essentially unchanged on cooling toward the critical glass transition temperature $\Tc$ of mode-coupling theory.  Spatial correlations between monomers on different chains, however, depend on temperature, albeit smoothly.  This implies that the glassy behavior of our model cannot result from static intra-chain or CM-CM correlations.  It must be related to inter-chain correlations at the monomer level.  Additionally, we study the dependence of inter-chain correlation functions on the position of the monomer along the chain backbone.  We find that this site-dependence can be well accounted for by a theory based on the polymer reference interaction site model (PRISM).  We also analyze triple correlations by means of the three-monomer structure factors for the melt and for the chains.  These structure factors are compared with the convolution approximation that factorizes them into a product of two-monomer structure factors.  For the chains this factorization works very well, indicating that chain connectivity does not introduce special triple correlations in our model.  For the melt deviations are more pronounced, particularly at wave vectors close to the maximum of the static structure factor.   
\end{abstract}


\maketitle

\section{Introduction}
\label{sec:intro_mct}

The microscopic understanding of the glass transition is a challenging problem in contemporary condensed matter physics \cite{LunkenheimerReview2000,Mezard:PhysicaA2002,BinderBaschnagelPaul2003}.  During the past decade the research in this field was strongly influenced by the mode-coupling theory (MCT) \cite{Goetze1999_review,Goetze_LesHouches}.  This theory suggests that a non-linear coupling between density fluctuations drives the slowing down of the structural relaxation when a li\-quid approaches its glass transition.  MCT predicts that there is a critical temperature $\Tc$, experimentally found to be above the calorimetric glass transition temperature $\Tg$, where the dynamics qualitatively changes.  For $T > \Tc$ the relaxation of the glass former is determined by the cooperative motion of the particles comprised in the nearest-neighbor shells (``cage-effect'').  In ``ideal MCT'', the simplest version of the theory, the mutual blocking of the particles in the cages leads to a complete structural arrest at $\Tc$.  This complete freezing is not observed experimentally, possibly with the exception of polydisperse hard-sphere-like colloidal suspensions \cite{vanMegen1995}.  When cooling the glass former toward $\Tg$ the structural relaxation time continuously increases, instead of diverging at $\Tc$.  Thus, alternative relaxation mechanisms must exist besides the cage effect and eventually become dominant for $T \lesssim \Tc$.  Within MCT the microscopic origin of these processes is, however, not well understood.

Despite this limitation of its range of validity the ideal MCT has been tested in numerous experiments \cite{Goetze1999_review} and computer simulations \cite{Goetze1999_review,kobreview1999,BinderBaschnagelPaul2003}.  Broadly speaking, the theoretical predictions were found to provide an adequate description of the relaxation dynamics above $\Tc$.  This success has stimulated extensions of the theory, originally developed for simple liquids, to molecules with orientational degrees of freedom \cite{TheisSciortino2000,FabbianLatzSchilling2000,ChongGoetze:PRE2002} and recently also to polymers \cite{ChongFuchs:PRL2002}. 

A distinguishing feature of the theory consists in establishing a quantitative link between the structure of a glass former and its dynamics.  This link may be exploited to predict the relaxation behavior provided the relevant static properties are available.  These properties involve the static structure factor and related quantities which must be determined with high precision over a large range of wave vectors.  Presumably due to this prerequisite a quantitative comparison of the predicted and measured dynamics has been attempted only for a few systems in the past, such as hard-sphere-like colloidal particles \cite{GoetzeSjoegren1991_colloids}, soft-sphere \cite{BarratLatz1990}, hard-sphere \cite{FoffiEtal:2003} or Lennard-Jones mixtures \cite{NaurothKob1997}, diatomic molecules \cite{WinklerEtal:PRE2000}, and models for {\em ortho}-terphenyl \cite{RinaldiSciortinoTartaglia2001,ChongSciortino:PRE} and $\mr{SiO_2}$ \cite{KobNaurothSciortino2002,SciortinoKob2001}.  These studies suggest that for $T \gtrsim \Tc$ MCT is a promising approach to a quantitative description of the structural relaxation for a large class of liquids comprising fragile and strong glass formers.

These findings motivate our present work.  An extension of MCT to non-entangled polymer melts \cite{ChongFuchs:PRL2002,ChongEtal:Polymer1} opens the possibility to attempt a quantitative comparison also for a polymeric glass former.  Here, we present the first step toward such a comparison for a simulated bead-spring model of a supercooled polymer melt \cite{BennemannPaulBinder1998,BennemannBaschnagelPaul1999_incoherent,BennemannPaulBaschnagel1999_Rouse,betaDynamics,alphaDynamics}.  We discuss various static structure factors, paying particular attention to the dependence of the structure on the position of a monomer along the polymer backbone and to correlation functions involving the center of the mass of the chains.  This information may be used to develop a tractable theory.  We will report on that and on the comparison with the simulations in a forthcoming article \cite{ChongEtal:Polymer1}.  A key aspect of this theory is that the short-range order of the monomers, as measured by the main peak of the collective static structure factor, strengthens with decreasing temperature.  The strengthening of the local packing provides the dominant mechanism causing structural arrest and glassy dynamics.  This mechanism, termed ``cage effect'' in simple liquids  \cite{Goetze1999_review,Goetze_LesHouches}, is also at the core of our theory for polymer melts.  In this article, we will thus pay special attention to structural correlations around the average spacing between monomers, and to the question of how chain connectivity affects them.

\section{Model and Simulation Technique}
\label{sec:model}

We study a bead-spring model of linear polymer chains \cite{BinderBaschnagelPaul2003,BennemannPaulBinder1998}.  All monomers interact via a truncated and shifted Lennard-Jones (LJ) potential
\begin{equation}
U_\mr{LJ}(r) = \left\{ \begin{array}{r@{,\quad}l}
4 \epsilon \left[ (\sigma/r)^{12} - (\sigma/r)^{6}\right] + C 
  & r < 2 r_\mr{min} \;,\\
0 & r \geq 2 r_\mr{min}\;.
\end{array} \right. 
\end{equation}
In the sequel, we will use LJ units ($\epsilon=1$, $\sigma=1$; furthermore, Boltzmann's constant $k_\mr{B}=1$ and the monomer mass $m=1$).  The constant $C=127/4096$ is chosen so that $U_\mr{LJ}$ vanishes continuously at $r = 2 r_\mr{min}$, $r_\mr{min} = 2^{1/6}$ being the minimum of the non-truncated potential.

In addition to $U_\mr{LJ}$, successive monomers along the polymer backbone interact via a FENE potential \cite{KremerGrest1990} 
\begin{equation}
U_\mr{FENE}(r) =
-\frac{k}{2}R_{0}^2 \ln \left[1-\left(\frac{r}{R_0}\right)^2 \right]
\end{equation}
with $R_0=1.5$ and $k=30$.  The superposition of the LJ- and FENE potentials leads to a steep effective bond potential with a sharp minimum at $\rbond = 0.9606$.

This choice of parameters has two important consequences. First, it prevents bonds from crossing each other.  This imposes topological constraints \cite{DoiEdwards} which ultimately lead to reptation-like dynamics in the limit of long chains \cite{KremerGrest1990,PuetzKremerGrest2000}.  Second, the bond potential locally distorts the regular arrangement of the monomers because it favors the inter-monomer distance $\rbond$ which is incompatible with $r_\mr{min}$.  

When cooling the melt from high $T$ the incompatibility of $\rbond$ and $r_\mr{min}$ impedes crystallization, but does not preclude it \cite{BuchholzPaulVarnikBinder:JCP2002,MeyerMuellerPlathe:JCP2001,MeyerMuellerPlathe:Macromolecules2002}.  For the melt to remain amorphous the chains should also be flexible.  This was pointed out in simulations of a semi-flexible bead-spring model in which large bond angles are energetically favored by a bending potential \cite{MeyerMuellerPlathe:JCP2001,MeyerMuellerPlathe:Macromolecules2002}.  The interplay of chain stiffness and excluded volume interactions suffices to induce crystallization from the melt.  Contrary to that, the chains of our model are flexible.  In the temperature range studied, the end-to-end distance ($R_\mr{e}^2 \simeq 12.3$) and the radius of gyration ($R_\mr{g}^2 \simeq 2.09$) are almost constant, and the collective static structure factor of the melt is typical of an amorphous material \cite{BennemannBaschnagelPaul1999_incoherent,betaDynamics}.

We analyze time series of isobaric simulations at the pressure $p=1$ \cite{BennemannPaulBinder1998,BuchholzPaulVarnikBinder:JCP2002}.  The polymer melt contains $n$ monodisperse chains of length $N=10$ in the volume $V$.  Depending on temperature (Nos\'e-Hoover thermostat) $n$ ranges between $n=100$ and $n=120$.  This corresponds to the following chain ($\rho$) and monomer densities ($\rho_\mr{m}$)
\begin{equation}
\begin{aligned}
0.091 \leq \rho & = \frac{n}{V} \leq 0.104  \;, \\
0.91 \leq \rho_\mr{m} & = \frac{nN}{V} \leq 1.04  \;.
\end{aligned}
\label{eq:defdensities}
\end{equation}

\section{Theoretical Background}
\label{sec:theory}

\subsection{Basic Notations}
\label{subsec:basic_notations}

Let $\vec{r}_i^a$ denote the position of the $a$th monomer in chain $i$ and $\vec{R}_i$ the position of the center of mass (CM) of chain $i$,
\begin{equation}
\vec{R}_i = \frac{1}{N} \sum_{a=1}^N \vec{r}_i^a \quad (i=1,\ldots,n) \;.
\label{eq:defCM}
\end{equation}
The knowledge of $\vec{r}_i^a$ and $\vec{R}_i$ allows us to define various density fluctuations for the wave vector $\vec{q}$ in reciprocal space: the density fluctuations of monomer $a$,
\begin{gather}
  \rho_a(\vec{q}) = \sum_{i=1}^n \exp\left[\mr{i}\vec{q}\cdot \vec{r}_i^a\right] \quad (a=1,\ldots,N) \;, \label{eq:def_rho_ai}  \\
\intertext{the density fluctuations of a tagged chain $i$ obtained by summing over all monomers of the chain,} 
\rho^\mr{p}_i(\vec{q}) = \sum_{a=1}^N \exp\left[\mr{i}\vec{q}\cdot \vec{r}_i^a\right]\;, \label{eq:def_rho_p}\\
\intertext{the density fluctuations created by all monomers of the melt,} 
  \rho_\mr{tot}(\vec{q}) = \sum_{i=1}^n \sum_{a=1}^N \exp\left[\mr{i}\vec{q}\cdot \vec{r}_i^a \right]\;, \label{eq:def_rho}\\
\intertext{and the polymer-density fluctuations related to the CM's of all chains} 
  \rho_\mr{C}(\vec{q}) = \sum_{i=1}^n
  \exp\left[\mr{i}\vec{q}\cdot \vec{R}_i\right] \;.  \label{eq:def_rho_C}
\end{gather}
Density-density correlation functions are well established means to describe the structure (and the dynamics) of a liquid \cite{HansenMcDonald}.  For a polymer melt we can derive various such two-point correlation functions from Eqs.~(\ref{eq:def_rho_ai}--\ref{eq:def_rho_C}).  They are introduced in the following section.

\subsection{Static Structure Factors}
\label{subsec:dens_corr}

The density-density correlations at the monomer level may be characterized by the monomer-monomer (or site-site) static structure factors 
\begin{equation}
S_{ab}(q) = \frac{1}{n} \left \langle  \rho_a(\vec{q})^* \rho_b(\vec{q})\right \rangle \label{eq:def_Sab} \;.
\end{equation}
Here, $\langle \cdot \rangle$ denotes the canonical average over all configurations of the melt.  Since the melt is spatially homogeneous and isotropic, the structure factors depend only on the modulus of the wave vector, $|\vec{q}|=q$.

We can split Eq.~\eqref{eq:def_Sab} into an intra-chain and an inter-chain part:
\begin{equation}
S_{ab}(q) = w_{ab}(q) + \rho h_{ab}(q) \;.
\label{eq:Sab_split}
\end{equation}
The intra-chain contribution is given by
\begin{equation}
w_{ab}(q) = \frac{1}{n} \Big \langle \sum_{i=1}^n \exp \bigl\{-\mr{i} 
\vec{q} \cdot [\vec{r}_i^a - \vec{r}_i^b]\bigr\} \Big \rangle \;,
\label{eq:def_wab}
\end{equation}
and the inter-chain contribution by
\begin{equation}
\rho h_{ab}(q) =\frac{1}{n} \Big \langle \sum_{i\neq j}^n \exp \bigl\{-\mr{i} \vec{q} \cdot [\vec{r}_i^a - \vec{r}_j^b]\bigr\} \Big \rangle \;.
  \label{eq:defSqABdistinct}
\end{equation}
These contributions reveal static correlations between monomers belonging to the same chain or to different chains (cf.\ \figref{fig:definitions2}).

\bfig
\rsfig[0.3]{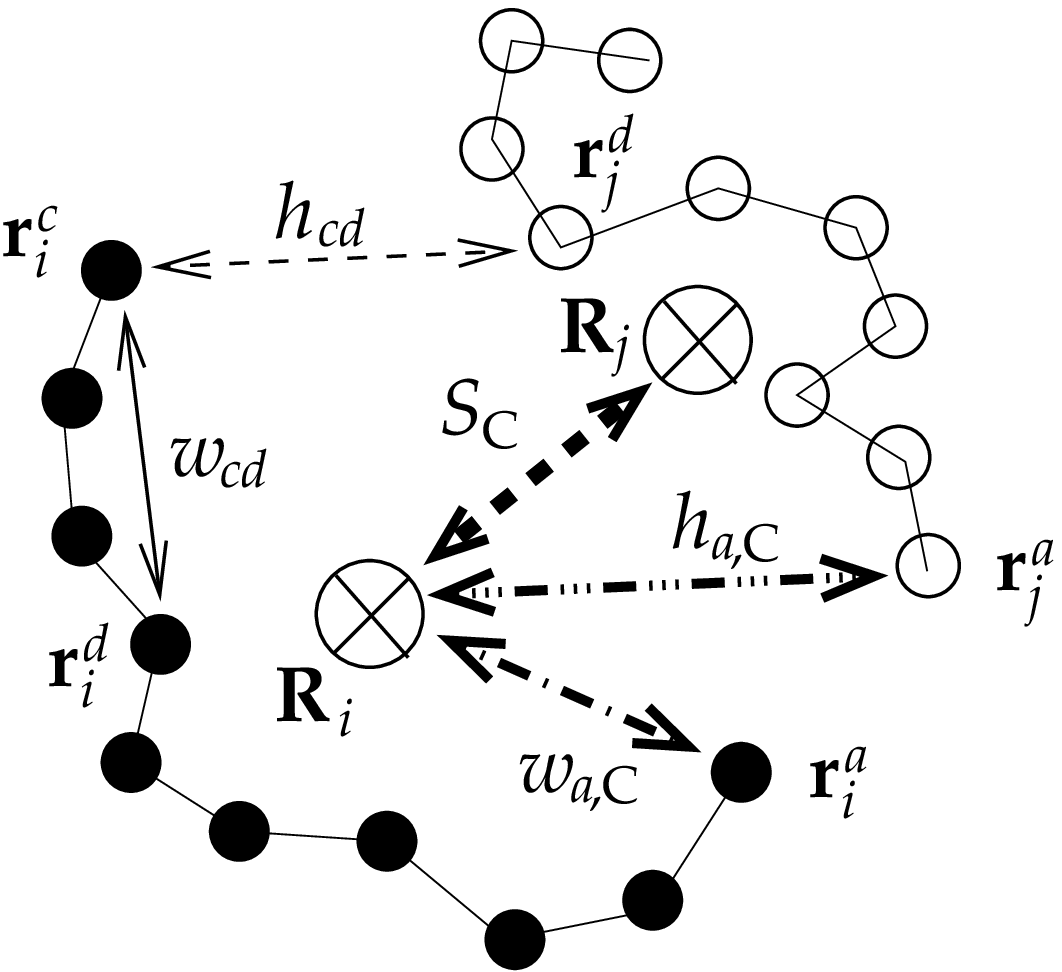}
\caption[]{
\label{fig:definitions2}
Schematic illustration of the correlation functions defined in Sect.~\ref{subsec:dens_corr}.  $\vec{r}_i^a$ is the position of the $a$th monomer of chain $i$ and $\vec{R}_i$ the position of the chain's center of mass.  $w_{cd}$ and $w_{a,\mr{C}}$ [Eqs.~(\ref{eq:def_wab},\ref{eq:defSqaCs})] denote intra-chain structure factors, $h_{cd}$ and $h_{a,\mr{C}}$ [Eqs.~(\ref{eq:defSqABdistinct},\ref{eq:defSqaCd})] inter-chain structure factors.  $S_\mathrm{C}$ [Eq.~\eqref{eq:defSqCM}] is the structure factor of the centers of mass.}
\efig

When averaging over all monomer pairs $(a, b)$ we obtain the collective static structure factor of the melt,
\begin{align}
  \label{eq:Sq}
  S(q) & = \frac{1}{nN} \left \langle  \rho_\mr{tot}(\vec{q})^* \rho_\mr{tot}(\vec{q}) \right \rangle = \frac{1}{N} \sum_{a,b=1}^N S_{ab}(q) \nonumber \\ 
  &\;= w(q) + \rho_\mr{m}h(q)\;,
\end{align}
where $\rho_\mr{m} = N \rho$ and
\begin{equation}
w(q) = \frac{1}{N} \left \langle  \rho^\mr{p}_i(\vec{q})^* \rho^\mr{p}_i(\vec{q}) \right \rangle = \frac{1}{N} \sum_{a,b=1}^N w_{ab}(q)
\label{eq:def_wq}
\end{equation}
denotes the static structure factor of a chain and
\begin{equation}
  h(q) = \frac{1}{N^2} \sum_{a,b=1}^N h_{ab}(q)
\label{eq:def_hq}
\end{equation}
is the Fourier transform of the site-averaged inter-molecular pair-correlation function \cite{HansenMcDonald}.

Usually, the averaged quantities $S(q)$ and $w(q)$ are used to characterize the structure of a polymer melt.  Contrary to that, we focus here on the monomer-resolved genera\-lizations $S_{ab}(q)$ and $w_{ab}(q)$.  The aim of our study is to understand to what extent specific monomer-monomer correlations deviate from the average behavior.  Since the structure factors $S_{ab}(q)$ and $w_{ab}(q)$ are important input quantities for the mode-coupling approach to glassy polymer dynamics \cite{ChongFuchs:PRL2002}, the comparison of $S_{ab}(q)$ and $w_{ab}(q)$ with their monomer-averaged counterparts can suggest suitable approximations and thus help developing a tractable theory \cite{ChongEtal:Polymer1}. 

In addition to density fluctuations of the monomers the spatial arrangement of the CM's and the coupling between the CM and the monomers can be analyzed.  We define the CM-CM structure factor (see \figref{fig:definitions2})
\begin{align}
S_\mr{C}(q) & = \frac{1}{n} \big \langle \rho_\mr{C}(\vec{q})^* \rho_\mr{C}(\vec{q}) \big \rangle \nonumber \\
& \;= \frac{1}{n} \Big \langle \sum_{i,j=1}^n \exp \bigl\{-\mr{i} \vec{q}
\cdot [\vec{R}_i- \vec{R}_j]\bigl\} \Big \rangle \nonumber \\
& \;= 1 + \rho h_\mr{C}(q) \;,
\label{eq:defSqCM}
\end{align}
which we split, in analogy to Eq.~\eqref{eq:Sq}, into self- (``1'') and distinct ($h_\mr{C}$) parts.  Formally, Eq.~\eqref{eq:defSqCM} is identical to that of simple liquids \cite{HansenMcDonald}.

Similarly, the coupling between a monomer and the CM's leads to monomer-polymer structure factors:
\begin{align}
S_{a,\mr{C}}(q) & = \frac{1}{n} \big \langle \rho_a(\vec{q})^* \rho_\mr{C}(\vec{q}) \big \rangle \nonumber \\
& \;= \frac{1}{n} \Big \langle \sum_{i,j=1}^n \exp \bigl\{-\mr{i} \vec{q} \cdot [\vec{r}_i^a - \vec{R}_j]\bigr\} \Big \rangle \;.
\label{eq:def_SaC}
\end{align}
$S_{a,\mr{C}}(q)$ is the Fourier transform of the (averaged) probability of finding a site $a$ at a distance $r$ from the center of a chain.  Following Eqs.~(\ref{eq:Sab_split}--\ref{eq:defSqABdistinct}) we separate again the intra-chain contribution,
\begin{equation}
\label{eq:defSqaCs}
w_{a,\mr{C}}(q) = \frac{1}{n} \Big \langle \sum_{i=1}^n
\exp \bigl\{-\mr{i} \vec{q} \cdot [\vec{r}_i^a - \vec{R}_i]\bigr\} \Big
\rangle \;,
\end{equation}
from the contribution involving different chains 
\begin{equation}
\label{eq:defSqaCd} 
\rho  h_{a,\mr{C}}(q) = \frac{1}{n} \Big \langle \sum_{i\neq j}^n
\exp \bigl\{-\mr{i} \vec{q} \cdot [\vec{r}_i^a - \vec{R}_j]\bigr\} \Big
\rangle \;.
\end{equation}
The correlations measured by Eqs.~(\ref{eq:defSqaCs},\ref{eq:defSqaCd}) are illustrated in \figref{fig:definitions2}. While the intra-chain structure factor $w_{a,\mr{C}}(q)$ can also be determined for a single polymer at infinite dilution \cite{SchaeferKrueger_JPhys1988}, the inter-chain $h_{a,\mr{C}}(q)$ describes how $a$-sites arrange around the CM of another polymer.  When summing over all monomers and defining
\begin{equation}
  \label{eq:defSqmC}
  w_\mr{m,C}(q) = \sum_{a=1}^N w_{a,\mr{C}}(q) \;, \quad
  h_\mr{m,C}(q) = \frac 1N \sum_{a=1}^N h_{a,\mr{C}}(q) \;.
\end{equation}
we obtain from Eqs.~(\ref{eq:def_SaC}--\ref{eq:defSqaCd})
\begin{equation}
\label{eq:defSqaC}
S_\mr{m,C}(q) = \sum_{a=1}^N S_{a,\mr{C}}(q) = w_\mr{m,C}(q) + \rho_\mr{m} h_\mr{m,C}(q) \;.
\end{equation}
These functions describe the averaged packing of sites around the center of mass of the same ($w_\mr{m,C}$) and of a different ($h_\mr{m,C}$) polymer, respectively.

\subsection{Three-Particle Structure Factors}
\label{subsec:s3}

$S(\vec{q})$ measures the spatial distribution of two monomers.  It depends on one wave vector $\vec{q}$.  Its generalization to a triple correlation function, the three-monomer structure factor $S_3(\vec{q},\vec{k})$, describes the (averaged) packing of a third monomer  which results from fixing the position of two monomers.  $S_3(\vec{q},\vec{k})$ depends on two wave vectors and is defined by
\begin{widetext}
\begin{equation}
\label{eq:def_s3}
S_3(\vec{q}, \vec{k}) = \frac{1}{n N} \big \langle \rho_\mr{tot}(\vec{-q}) \rho_\mr{tot}(\vec{k}) \rho_\mr{tot}(\vec{q} - \vec{k}) \big \rangle = \frac{1}{n N} \Big \langle \sum_{i,j,l=1}^{n}\sum_{a,b,c=1}^{N} 
\exp\bigl\{\mr{i}[-\vec{q}\cdot \vec{r}_i^a + \vec{k}\cdot \vec{r}_j^b + (\vec{q}-\vec{k})\cdot \vec{r}_l^c]\bigr\} \Big \rangle \;,
\end{equation}
\end{widetext}
where the vector $\vec{q} - \vec{k}=\vec{p}$ is the third side of a triangle forming an angle $\varphi$ between $\vec{q}$ and $\vec{k}$. 
The angle $\varphi$ is given by
\begin{equation}
\label{phi_qk}
\cos\varphi = \frac{q^2 + k^2 - p^2}{2 q k}\;.
\end{equation}


$S_3(\vec{q}, \vec{k})$ is the Fourier transform of the three particle distribution function, which gives the average density of (other) segments at a space point if the positions of two segments are fixed. It is related to a direct correlation function $c_3(\vec{q},\vec{k})$ by the triplet Ornstein-Zernike equation \cite{BarratHansenPastore1987}
\begin{equation}
\label{eq:S_3_and+c_3}
S_3(\vec{q},\vec{k}) = S(q)S(k)S(|\vec{q}-\vec{k}|)\left[1+ \rho_\mr{m}^2 c_3(\vec{q},\vec{k}) \right] \;.
\end{equation}
Often, when considering $S_{3}$ for wavevectors corresponding to the average particle distance, it is assumed that there are no three-body correlations which are not contained in the product of two-particle correlation functions.  This so-called convolution approximation~\cite{BarratHansenPastore1987,JacksonFeenberg1962} --note that it differs from the Kirkwood superposition approximation which becomes valid for large distances \cite{HansenMcDonald}-- implies that the triplet direct correlation function vanishes, $c_3(\vec{q},\vec{k}) \equiv 0$.  Here, we want to test this approximation for $S_3(\vec{q},\vec{k})$ and for the polymer three-monomer structure factor $w_3(\vec{q},\vec{k})$.  $w_3(\vec{q}, \vec{k})$ is defined analogously to $S_3$ with $\rho_\mr{tot}$ replaced by $\rho^\mr{p}_i$ in Eq.~\eqref{eq:def_s3}, that is, only the term $i=j=l$ of the sum in Eq.~\eqref{eq:def_s3} is taken into account.

For a homogeneous and isotropic system $S_3$ depends on the moduli of the three wave vectors only, $S_3(\vec{q},\vec{k})=S_3(q,k,p=|\vec{q}-\vec{k}|)$.  To determine $S_3(q,k,p)$ we utilized a method closely related to the one proposed in Ref.~\cite{SciortinoKob2001}.  The triple of moduli $(q,k,p)$ satisfying 
Eq.~\eqref{phi_qk}
is discretized in bins of width $\Delta q = \Delta k = \Delta p = 0.2$.  In each bin 100 vector tuples $\{(\vec{q}, \vec{k}) \; \big | \; |\vec{q}|=q, |\vec{k}|=k, |\vec{p}| = p)\}$ are chosen at random for each configuration and $S(q,k,p)$ is calculated as the average over this set of vectors and all configurations.  Data were accumulated over 1155 configurations at $T=0.47$ \cite{Comment_on_S3}.

\section{Results on Two-Point Correlation Functions}
\label{sec:results_statics}

\subsection{Static Properties at the Monomer Level: Site-averaged Quantities}
\label{subsec:stat_monomers_average}

We discuss the static structure factor of the melt and the corresponding self- and distinct parts [Eqs.~(\ref{eq:Sab_split}--\ref{eq:def_hq})].  For $T<1$ and $q \leq 20$, results for $S(q)$ and $w(q)$ have been presented previously \cite{BinderBaschnagelPaul2003,BennemannBaschnagelPaul1999_incoherent,betaDynamics,alphaDynamics}.  Here, we extend the analyses up to $q=50$ and considerably improve the statistics (averages over more than 1000 configurations).  This effort was necessary to use the static quantities in MCT calculations which require a large $q$-range and good statistics. 

\subsubsection{Static Structure Factor of the Melt}
\label{subsubsec:statics_melt}

Figure~\ref{fig:S_q_melt_allT} shows $S(q)$ for temperatures above the critical temperature of MCT ($\Tc \simeq 0.45$).  In this $T$-interval the structure of the melt is typical of a disordered, dense system.  Due to the weak compressibility of the melt $S(q)$ is small in the limit $q \rightarrow 0$.  As $q$ increases, $S(q)$ increases toward a maximum which occurs around $q_\mr{max} \simeq 7.15$ in our model.  This $q$-value corresponds to the length scale of the monomer diameter ($= 1$).  Thus, the dominant contribution to $S(q_\mr{max})$ comes from the amorphous packing in the nearest neighbor shell around a monomer.  On cooling, Fig.~\ref{fig:S_q_melt_allT} indicates that no long-range structural correlations develop in the melt.  Only its density increases and the packing becomes tighter.  Two features reflect these changes: the amplitude of the peak $S(q_\mr{max})$ grows and its position $q_\mr{max}$ shifts to larger values.  Close to $q_\mr{max}$, the dependence of $S(q)$ on $T$ is most pronounced.

\bfig 
\rsfig{S_q_melt_allT_new.eps}
\caption[]{\label{fig:S_q_melt_allT}
Collective static structure factor $S(q)$ of the melt versus the modulus of the wave vector $\vec{q}$ [Eq.~\eqref{eq:Sq}].  The temperatures shown are: $T=1$ (dashed line), 0.7, 0.65, 0.6, 0.55, 0.52, 0.5, 0.48, 0.47, and 0.46 (solid line).  Inset: First maximum of $S(q)$, $S(q_\mr{max})$, versus $T$ ($q_\mr{max} \approx 7.15$).  The dashed horizontal line indicates the Hansen-Verlet criterion for the glass transition of hard spheres within the ideal MCT ($S(q_\mr{max}) \approx 3.54$) \cite{FuchsSchweizer:JPCM2002}.} 
\efig

The increase of $S(q_\mr{max})$ with decreasing $T$ may be related to the (empirical) Hansen-Verlet freezing criterion \cite{HansenVerlet}.  This criterion states that a liquid will condense to a solid phase when $S(q_\mr{max})$ exceeds a threshold value.  For crystallization this threshold is $S(q_\mr{max}) \approx 2.85$ \cite{Comment_on_freezing_criteria}.  For the glass transition of hard spheres within the ideal MCT one finds $S(q_\mr{max}) \approx 3.54$ \cite{FuchsSchweizer:JPCM2002}.  The latter value agrees very well with our simulation result for $S(q_\mr{max})$ at $T=0.46$ (see inset of \figref{fig:S_q_melt_allT}), indicating that the melt is close to the $\Tc$ of the ideal MCT.  Indeed, our previous studies \cite{BinderBaschnagelPaul2003,BennemannBaschnagelPaul1999_incoherent,BennemannPaulBaschnagel1999_Rouse,betaDynamics,alphaDynamics} suggest $\Tc \simeq 0.45$.
 
This interpretation of the structure factor only involves packing arguments which could also be put forward for atomic liquids.  To obtain a better insight into the role of chain connectivity we split $S(q)$ into intra-chain (self) and inter-chain (distinct) contributions [Eqs.~(\ref{eq:Sab_split}--\ref{eq:def_hq})].  We discuss these contributions in the next two sections.

\subsubsection{Intra-chain Structure}
\label{subsubsec:statics_intra}

Figure~\ref{fig:w_q+approx_all_T} shows the intra-chain contribution to $S(q)$, the structure factor $w(q)$ of a polymer, at $T=0.46$ and $T=1$.  Both temperatures yield almost identical results for all $q$. So, data at intermediate $T$ are not included in the figure.  

\bfig 
\rsfig{w_q+approx_all_T.eps}
\caption[]{
\label{fig:w_q+approx_all_T}
Polymer static structure factor $w(q)$ [Eq.~\eqref{eq:def_wq}] at the lowest and the highest temperature, $T=0.46$ (solid line) and $T=1.0$ (dashed line), respectively.  The simulation data are compared to the Debye formula (Eq.~\eqref{eq:Debye}, dotted line) and to the large-$q$ approximation Eq.~\eqref{eq:wq_approx} ($\rbond=0.9609$, dash-dotted line).  The inset magnifies of the small-$q$ behavior of $w(q)$.}
\efig

The independence of $T$ may be rationalized in the following way.  If $q\rightarrow \infty$, local rapid variations will determine the behavior of $w(q)$.  In this limit, we can approximate Eq.~\eqref{eq:def_wq} by
\begin{equation}
  \begin{split}
  w(q) 
    &\approx 1 + \frac2N (N-1)\, w_{a\,a+1}(q) \quad (\mbox{$q$ large}) \;,
  \end{split}
  \label{eq:w_high_q}
\end{equation}
where we also assumed that the nearest-neighbor contribution $w_{a\,a+1}(q)$ is independent of $a$.  This assumption is well justified for our model, as a glance at Fig.~\ref{fig:w_ab_T0.47_1.1-1.10+inner} shows.  $w_{a\,a+1}(q)$ can be evaluated by exploiting that $|\vec{r}^a - \vec{r}^{a+1}| \approx \rbond$ due the stiff bond potential of our model.  Thus,
\begin{align}
\label{eq:w_a_a+1}
w_{a\,a+1}(q) 
&= \bigg \langle \frac{\sin(q |\vec{r}^a - \vec{r}^{a+1}|)}{q |\vec{r}^a - \vec{r}^{a+1}|} \bigg \rangle \approx \frac{\sin(q \rbond)}{q\rbond} \;.
\end{align}
To lowest order, we therefore expect
\begin{equation}
  \label{eq:wq_approx}
  w(q) \approx 1 + \frac2N (N-1)\, \frac{\sin(q \rbond)}{q \rbond}  \quad (\mbox{$q$ large}) \;.
\end{equation}

For small $q$-values polymer physics suggests that the Debye function $w_\mr{D}(q)$ \cite{DoiEdwards,StroblPolymerPhysics} provides a reasonable description of $w(q)$,
\begin{equation}
w_\mr{D}(q) = Nf_\mr{D}(q^2 R^2_\mr{g})\,, \;\; f_\mr{D}(x) = \frac2{x^2}\left[ \mr{e}^{-x} + x -1\right]\;.
  \label{eq:Debye}
\end{equation}
The Debye theory assumes a Gaussian distribution for all distances along the backbone of the chain.  This assumption correctly reflects the random-walk-like structure of (long) polymers on intermediate and large length scales in the melt.  However, at large $q$ where the precise form of the interaction potential matters, it cannot apply.

The approximations \eqref{eq:wq_approx} and \eqref{eq:Debye} are compared to the simulation data in \figref{fig:w_q+approx_all_T}.  For small $q$, the scattering is determined by the size of the polymer only.  In this limit, the Debye theory reproduces the exact result $w_\mr{D}(q) = N(1 - q^2R^2_\mr{g}/3)$.  So, it has to coincide with the measured $w(q)$ if $q < 1/R_\mr{g}\simeq 0.69$ (see inset of \figref{fig:w_q+approx_all_T}).  Furthermore, the theory and the simulation should also agree for $1/R_\mr{g} \ll q \ll 1/\rbond$.  This $q$-range probes the random-walk-like internal structure of a polymer, where both $w(q)$ and $w_\mr{D}(q)$ scale as $\sim q^{-2}$.  However, since our chains are short, the length scales $\rbond$ and $R_\mr{g}$ are not sufficiently separated for this behavior to be observed.  Instead, $w(q)$ crosses over to regular oscillations which are compatible with Eq.~\eqref{eq:wq_approx} for $q \gtrsim 8$.

Figure~\ref{fig:w_q+approx_all_T} shows that the superposition of Eqs.~\eqref{eq:wq_approx} and \eqref{eq:Debye} approximately describes the simulation data.  Thus, in our model the main features of $w(q)$ are determined by two length scales, $\rbond$ and $R_\mr{g}$.  Since $R_\mr{g}$ depends only weakly on $T$ and $\rbond$ is independent of $T$, the almost perfect agreement of the structure factors for $T=0.46$ and $T=1$ can be understood.

\subsubsection{Inter-chain Structure}
\label{subsubsec:statics_inter}

If the intra-chain contribution to $S(q)$ is independent of $T$, the temperature dependence of $S(q)$ must result from inter-chain correlations.  Figure~\ref{fig:S_dist_q_allT} supports this expectation.  The distinct part $\rho_\mathrm{m}h(q)$ [Eq.~\eqref{eq:Sq}] exhibits liquid-like oscillations whose extrema become more pronounced on cooling.  This trend is especially visible around $q_\mr{max}$ ($\simeq 7.15$), suggesting that the glassy behavior of our model is driven by $h(q)$ only.  That is, by nearest neighbors which are not bonded to one another.  This finding is not unreasonable.  As the distance between successive monomers along the backbone is almost fixed, only non-bonded neighbors can pack more tightly and reinforce the cage with decreasing $T$.

\bfig
\rsfig{S_dist_q_allT_new.eps}
\caption[]{\label{fig:S_dist_q_allT}
Distinct contribution to the static structure factor $\rho_\mr{m} h(q)$ [Eq.~\eqref{eq:Sq}] at $T=0.46$, $0.65$ and $1.0$.  Inset: Comparison of $\rho_\mr{m} h(q)$ with $S(q)-1$ and $-w_\mr{D}(q)$ at $T=0.46$ (see text for further details).}
\efig

We support this interpretation by the following argument:  
If our system was a simple liquid, we would have $w(q)=1$, and $\rho_\mr{m}h(q)=S(q)-1$ would be exact.  The inset of \figref{fig:S_dist_q_allT} shows that this simple-liquid-like approximation represents a good description for $q \gtrsim 6$.  For smaller $q$, deviations are found.  Here, $S(q)$ becomes vanishingly small ($S(q) \sim 10^{-2}$, see \figref{fig:S_q_melt_allT}) and polymer-specific effects, i.e., the correlation hole \cite{DeGennes}, determine the distinct part: $\rho_\mr{m}h(q) \approx - w(q)$.  The correlation hole implies that the probability of finding monomers of other polymers inside the volume $V_\text{c}$ occupied by a chain is decreased.  This effect arises because the probability of finding the monomers of a chain in its own volume $V_\text{c}$ is enhanced, and intra- and inter-moleculer correlations have to compensate each other to render the melt incompressible \cite{DeGennes,SchweizerCurro1997}.

\subsection{Static Properties at the Monomer Level: Site-resolved Quantities}
\label{subsec:stat_monomers_resolved}

\subsubsection{Direct Correlation Function}
\label{subsubsec:direct_correlation}

For simple liquids the direct correlation function is usually introduced through the Ornstein-Zernike equation \cite{HansenMcDonald}.  For molecular liquids Chandler and Andersen \cite{ChandlerAndersen1972} extended the Ornstein-Zernike approach to reflect the contribution from the intra-molecular correlations $w_{ab}(q)$.  The resulting generalized site-site Ornstein-Zernike equation --also referred to as ``Reference Interaction Site Model'' (RISM)-- is given by \cite{HansenMcDonald}
\begin{equation}
h_{ab}(q) = \sum_{x,y=1}^{N}
w_{ax}(q) \, c_{xy}(q) \,
\big [w_{yb}(q) + \rho h_{yb}(q)\big] \;.
\label{eq:RISM}
\end{equation}
Here, $c_{ab}(q)$ is the direct correlation function between the sites $a$ and $b$.  Inserting Eq.~\eqref{eq:RISM} into Eq.~\eqref{eq:Sab_split} we obtain
\begin{equation}
\rho c_{ab}(q) = \big [w_{ab}^{-1}(q) - S_{ab}^{-1}(q)\big ] \; ,
\label{eq:def_cab}
\end{equation}
where $X_{ab}^{-1}(q)$ denotes the $(a,b)$-element of the inverse of the matrix $\matr{X}(q)$.

The difficulty in dealing with site-site correlation functions arises from the dependence on the indices $(a,b)$.  This leads to $O(N^{2})$ coupled equations which cannot be handled for large $N$.  However, one can argue that, for long chains, end effects on {\em inter-polymer} correlations should be small, suggesting to treat all sites of a homopolymer equivalently.  (This simplification becomes exact for ring homopolymers.)  This {\em equivalent-site approximation} is usually invoked for $c_{ab}(q)$, i.e.,
\begin{equation}
c_{ab}(q)=c(q) \quad (\mbox{equivalent-site approximation})
\;.
\label{eq:PRISM-c}
\end{equation}
Equation~\eqref{eq:PRISM-c} represents the principal idea of the polymer RISM (PRISM) theory developed by Schweizer and coworkers \cite{SchweizerCurro1997}.

Inserting the assumption \eqref{eq:PRISM-c} into Eq.~\eqref{eq:RISM} we obtain from Eq.~\eqref{eq:def_hq} the so-called PRISM equation for the site-averaged pair-correlation function
\begin{equation}
h(q) = w(q) \, c(q) \, \big [w(q) + \rho_\mr{m} h(q) \big] 
\;. 
\label{eq:PRISM}
\end{equation}
Equations~(\ref{eq:Sq},\ref{eq:PRISM}) provide an expression for $c(q)$ in terms of $w(q)$ and $S(q)$:
\begin{equation}
\rho_\mr{m}c(q) = \frac{1}{w(q)} - \frac{1}{S(q)} \;.
\label{eq:PRISM-S1}
\end{equation}

\bfig 
\rsfig{site_aa_dcf_T0.47.eps}
\rsfig{site_ab_dcf_T0.47.eps}
\caption[]{\label{fig:c-site}
Examination of the equivalent-site approximation Eq.~(\ref{eq:PRISM-c}) at $T = 0.47$.  The solid lines in both panels denote $c(q)$ determined from the simulation results for $S(q)$ and $w(q)$ via Eq.~(\ref{eq:PRISM-S1}).  The dashed lines represent $c_{11}(q)$ (upper panel) and $c_{15}(q)$ (lower panel), the dotted lines $c_{22}(q)$, $c_{33}(q)$, $c_{44}(q)$, $c_{55}(q)$ (upper panel) and $c_{25}(q)$, $c_{35}(q)$, $c_{45}(q)$, $c_{55}(q)$ (lower panel).  The site-site direct correlation functions $c_{ab}(q)$ are calculated from the simulation results for $S_{ab}(q)$ and $w_{ab}(q)$ via Eq.~(\ref{eq:def_cab}).  The insets magnify the region close to $q_\mr{max}$.}
\efig

Figure~\ref{fig:c-site} examines the validity of the equivalent-site approximation by comparing Eq.~\eqref{eq:def_cab} and Eq.~\eqref{eq:PRISM-S1}.  Apparently, the approximation is well satisfied, except for functions involving the chain ends ($a=1$ or $a=N$).  Here, we find slight deviations close to $q_\mr{max}$ and more pronounced ones for $q \lesssim 5$.  From the point of view of MCT, the important wave-vector regime is around $q_\mr{max}$.  Thus, \figref{fig:c-site} suggests that, for our model, MCT equations for the dynamics of the melt can be derived by assuming Eq.~\eqref{eq:PRISM-c} without introducing a large error \cite{ChongEtal:Polymer1}.  In the following, we want to use Eq.~\eqref{eq:PRISM-c} to interpret the inter-molecular site-site-correlations of our model.

\subsubsection{Intra-chain Structure}
\label{subsubsec:statics_intra_ab}

If $w(q)$ does not depend on temperature, we may expect the same feature for all components $w_{ab}(q)$.  Indeed, we find that $w_{ab}(q)$ is (almost) independent of $T$.  Thus, we concentrate on one temperature in the following. 

\bfig
\rsfig{w_ab_T0.47_1.1-1.10+inner.eps}
\caption[]{
\label{fig:w_ab_T0.47_1.1-1.10+inner}
Static structure factor $w_{ab}(q)$ of the monomer pair $(a,b)$ at $T=0.47$.  $a$ and $b$ are monomers of the same chain.  Note that $w_{ab}(q)$ depends on $|a-b|$ only.  The simulation data for $|a-b|=1$ ($a=1,b=2$ and $a=5,b=6$) are compared with Eq.~\eqref{eq:w_a_a+1} (circles).  For separations $|a-b|=1,\ldots,5$ the Gaussian approximation Eq.~\eqref{eq:wq_Gauss} is also shown (dotted lines).} 
\efig

Figure~\ref{fig:w_ab_T0.47_1.1-1.10+inner} depicts $w_{ab}(q)$  at $T=0.47$ for various monomer pairs $(a,b)$.  We find that the explicit dependence of $w_{ab}(q)$ on the site indices is negligible for all wave vectors.  To a very good approximation, $w_{ab}(q)$ only depends on the distance $|a-b|$, a feature also found for a model of freely-jointed hard-sphere chains \cite{YethirajEtal:JCP1990} and for the Gaussian approximation 
\begin{equation}
  \label{eq:wq_Gauss}
  w_{ab}^\mr{G}(q) = \exp\big [-q^2|a-b|\lstat^2/6 \big ]\; .
\end{equation}
To compare Eq.~\eqref{eq:wq_Gauss} with the simulation data we identify the statistical segment length $\lstat$ with $\lstat^2 = R^2_\mr{e}/N = 6R^2_\mr{g}/N$ \cite{DoiEdwards}.  This assumption is valid for (long) chains in the melt.  Figure~\ref{fig:w_ab_T0.47_1.1-1.10+inner} shows that Eq.~\eqref{eq:wq_Gauss} only provides a good approximation for small $q$.  (The same result was also found when comparing $w(q)$ and $w_\mr{D}(q)$.)  With increasing $q$, $w_{ab}(q)$ decays faster than $w_{ab}^\mr{G}(q)$, if $|a-b|>1$, and becomes negative before approaching 0 from below.  This undershoot shifts to larger $q$ and increases in amplitude, as $|a-b|$ decreases toward 1.  Adjacent monomers along the backbone of the chain ($|a-b|=1$) exhibit long-range oscillations which are well described by Eq.~\eqref{eq:w_a_a+1}.

\subsubsection{Inter-chain Structure}
\label{subsubsec:statics_inter_ab}

\bfig
\rsfig{S_dist_ab_q_T0.47_new2.eps}
\caption[]{\label{fig:S_dist_ab_q_T0.47_new}
Inter-chain static structure factor, $\rho h_{ab}(q)$, at $T=0.47$ for different  pairs $(a,b)$ [Eq.~\eqref{eq:defSqABdistinct}].  $\rho h_{ab}(q)$ depends on $(a,b)$ at $q_\mr{max} \simeq 7.15$ and for $q \lesssim 5.5$.  The correlation of two chain ends ($a=b=1$) behaves differently in comparison to all other curves.   The average over all monomers, $\rho h(q)$, is also included.} 
\efig

Figure~\ref{fig:S_dist_ab_q_T0.47_new} shows the site-resolved pair correlations $h_{ab}(q)$ [Eqs.~(\ref{eq:Sab_split},\ref{eq:defSqABdistinct})] at $T=0.47$.  For $q \gtrsim 15$, $h_{ab}(q)$ is independent of the monomer index and coincides with the site-averaged $h(q)$.  Contrary to that, we find a dependence on $(a,b)$ around $q_\mr{max}$ and particularly for $q \lesssim 5.5$.  While pairs resulting from inner sites ($ 1 < a,b < N$), except $h_{22}(q)$, are still very close to each other, differences occur for correlations comprising an end monomer.  For $q \approx 4$, $h_{ab}(q)$ exhibits a shoulder.  If an end monomer is involved, the amplitude of the shoulder decreases and is smallest for the correlation of two chain ends ($h_{11}$).

Qualitatively, the site-dependence around $q_\mr{max}$ may be explained by the following argument.  As the analysis in Sect.~\ref{subsubsec:statics_monCM} will show, a middle segment is buried deeply in the polymer coil and is closer to the CM of its chain than is an end segment.  Thus, it is not surprising that end monomers have stronger local inter-molecular correlations, viz.\ that $\rho h_{11}(q)$ deviates more from zero around $q_{\rm max}$ than does $\rho h_{aa}(q)$ for middle segments.

\bfig 
\rsfig{h_aa_and_PRISM_T0.47.eps}
\rsfig{h_ab_and_PRISM_T0.47.eps}
\caption[]{\label{fig:h-site}
Comparison of the site-site inter-molecular pair-correlation functions $h_{ab}(q)$ determined from the simulation data at $T = 0.47$ (circles) and from the PRISM theory (grey solid lines) \cite{Comment_on_PRISM}.  Some curves are shifted vertically for clarity.  Note that $h_{21}(q)=h_{2\,10}(q)$ in the PRISM theory (see text for details).} 
\efig

A quantitative explanation of the site-dependence of $h_{ab}(q)$ may be obtained by PRISM theory.  From Eqs.~(\ref{eq:Sab_split}) and (\ref{eq:RISM}) it follows that
\begin{equation}
\begin{aligned}
S_{ab}(q) &=
\bigl[ \,
\{ \unitmatr - \rho \matr{w}(q) \matr{c}(q) \}^{-1} \, \matr{w}(q)
\, \bigr]_{ab} \; , \\
h_{ab}(q) &=
\bigl[ \,
\matr{w}(q) \matr{c}(q) \, 
\{ \unitmatr - \rho \matr{w}(q) \matr{c}(q) \}^{-1} \, \matr{w}(q)
\, \bigr]_{ab} \; ,
\end{aligned}
\label{eq:PRISM-sh}
\end{equation}
where $\unitmatr$ denotes the unit matrix. Thus, even with the assumption $c_{ab}(q) = c(q)$, a site-dependence of $S_{ab}(q)$ and $h_{ab}(q)$ results from chain connectivity due to the matrix structure of $w_{ab}(q)$.  E.g., for $h_{ab}(q)$ one finds [from Eqs.~(\ref{eq:Sq},\ref{eq:RISM},\ref{eq:PRISM})]
\begin{equation}
h_{ab}(q) = \frac{h(q)}{w(q)^2} \bigg [\sum_{x=1}^N w_{ax}(q) \bigg ]
\bigg [\sum_{y=1}^N w_{by}(q) \bigg ] \;.
\label{eq:PRISM-h}
\end{equation}
Figure~\ref{fig:h-site} compares theory and simulation for some representative pair-correlation functions $h_{ab}(q)$.  We find that $h_{ab}(q)$ is well described by Eq.~\eqref{eq:PRISM-h}.  This explains why the correlation hole in $h_{ab}(q)$ for end segments is slightly more narrow than for middle segments.  Furthermore, according to Eq.~(\ref{eq:PRISM-h}) $h_{ab}(q)$ should exhibit the symmetry: $h_{a,b}(q)= h_{a,N-b+1}(q)$.  This symmetry is tested in Fig.~\ref{fig:h-site} for $a=2$, $b=1$ and found to be well borne out.

A corresponding analysis for $S_{ab}(q)$ (not shown) finds the same agreement between the PRISM theory and the simulation data.  These results indicate that a complete description of the static structure of our polymer melt can be achieved using only the site-independent inter-chain direct correlation function $c(q)$ and the matrix of the single-chain structure factors $w_{ab}(q)$.

\subsection{Static Structure involving the Center of Mass}
\label{subsec:stat_cm}

\subsubsection{Structure Factor of the Center of Mass}
\label{subsubsec:statics_cm}
At low temperature, the motion of the CM slows down similarly to that of the monomers \cite{BennemannPaulBaschnagel1999_Rouse}.  For the monomers, we interpreted this behavior as a consequence of the tighter packing in the nearest-neighbor shells.  If a tighter packing of the polymer coils and concomitant ``coil-caging'' was responsible for the sluggish dynamics of the chains, one could expect to find the signature of a stronger packing in the CM-CM structure factor $S_\mr{C}$.  Alternatively, polymer MCT \cite{ChongFuchs:PRL2002,ChongEtal:Polymer1} suggests that the slowing down of the CM originates from the segment dynamics which ``enslaves'' the CM motion.  For this view to apply, little or no variation with $T$ of the CM correlations is required, as the inter-molecular segment correlations, $\rho_\mr{m}h(q)$ (Fig.~\ref{fig:S_dist_q_allT}), drive the (segment) caging.

Recently, the spatial correlations between the CM's have been addressed in several studies \cite{Guenza:PRL2002,Guenza:Macromolecules2002,BohuisEtal:JCP2001,BohuisEtal:PRE2001,KrakoviackEtAl:EPL2002}.  References~\cite{Guenza:PRL2002,Guenza:Macromolecules2002} point out that deviations of the CM motion from free diffusion, observed for displacements smaller than the chain size already at high $T$, could be caused by inter-molecular interactions between the centers of mass.  Approximately, these interactions are given by the potential of mean force \cite{HansenMcDonald}, i.e., by the CM-CM pair distribution function.  

This function was also discussed in the context of devising efficient coarse-grained simulation models for semi-dilute polymer solutions and polymer-colloid mixtures \cite{BohuisEtal:JCP2001,BohuisEtal:PRE2001}.  In these studies, the attempt is made to represent the polymer coils as soft, penetrating spheres.  The spheres interact via an effective pair potential derived from the CM-CM distribution function.  For this distribution function Ref.~\cite{KrakoviackEtAl:EPL2002} suggests a PRISM-approximation which relates $S_\mr{C}(q)$ to the structure factors of the monomers.  In Appendix~\ref{app:krakoviack} we sketch the main ideas of this approach and discuss the validity of the underlying assumptions for our polymer melt.  Here, we only compare the result of the calculation,
\begin{equation}
\label{eq:krak_approx2_in_text}
S_\mr{C}(q) \approx 1 + \frac{1}{N}\frac{w_\mr{m,C}(q)^2}
{w(q)^2}\, \rho_\mr{m} h(q) \;,
\end{equation}
with our simulation data [$w_\mr{m,C}(q)$ is defined in Eq.~\eqref{eq:defSqmC}].  

\bfig 
\rsfig{S_chain_cm_all_T+krakoviack_approx.eps}
\caption[]{
\label{fig:S_chain_cm_all_T+krakoviack_approx}
Static structure factor of the CM, $S_\mr{C}(q)$, at all simulated temperatures, i.e., $T=0.46,\ldots, 1$ (solid lines).  The dashed line shows Eq.~\eqref{eq:krak_approx2_in_text} calculated from the simulation data for $w_\mr{m,C}(q)$, $w(q)$ and $h(q)$ at $T=1$.  The dotted horizontal line indicates the limit of $S_\mr{C}(q)$ for $q\rightarrow 0$ [Eq.~\eqref{eq:SC_limit_q_to_0}]. $k_\mr{B}T \rho_\mr{m} \kappa_T$ was read off from the small-$q$ behavior of $S(q)$ (see \figref{fig:S_q_melt_allT}).  The dotted vertical line shows $1/R_\mr{g} \simeq 0.69$ ($R_\mr{g} \simeq 1.45$).}
\efig

Figure~\ref{fig:S_chain_cm_all_T+krakoviack_approx} shows $S_\mr{C}(q)$ for all investigated temperatures, together with Eq.~\eqref{eq:krak_approx2_in_text}.  Starting from a small value ($\kappa_T=$ isothermal compressibility),
\begin{equation}
\lim_{q\rightarrow 0} S_\mr{C}(q) = \frac{\langle n^2 
\rangle- \langle n \rangle^2}{\langle n \rangle} = \frac{1}{N} k_\mr{B} T \rho_\mr{m} \kappa_T \;,
\label{eq:SC_limit_q_to_0}
\end{equation}
the structure factor increases toward a small peak before it approaches the ideal gas value 1 without any further oscillations.  Although this peak is indicative of some preferred distance between the chains, the effect is very weak.  Furthermore, $S_\mr{C}(q)$ is independent of $T$.  This implies that the sluggish dynamics of the CM is not related to a tighter packing of the chains at low $T$.  Rather it should be interpreted as a consequence of the slowing down of the monomer motion, which, due to chain connectivity, entails the glassy behavior of the CM. 

As observed in Ref.~\cite{KrakoviackEtAl:EPL2002}, we also find that the PRISM-approximation~\eqref{eq:krak_approx2_in_text} provides a good description of the simulation data, except for $q$ close to the peak position, where $S_\mr{C}(q)$ is slightly underestimated.  Thus, we may use Eq.~\eqref{eq:krak_approx2_in_text} to interpret the finding that $S_\mr{C}(q)$ does not change on cooling.  Equation~\eqref{eq:krak_approx2_in_text} contains the prefactor $w_\mr{m,C}(q)^2/w(q)^2$, which depends on intra-chain correlations only, and is thus independent of $T$ for our model.
\bfig
\rsfig{S_sC_over_w_squared_T046+T1.0+Gauss.eps}
\caption[]{
\label{fig:S_sC_over_w_squared_T046+T1.0+Gauss}
$w_\mr{m,C}(q)^2/w(q)^2$ versus $q$ at $T=0.46$ (solid line) and $T=1$ (dashed line).  As $w(q)$ does not depend on temperature (see Fig.~\ref{fig:w_q+approx_all_T}), the figure indicates that the monomer-CM correlation $w_\mr{m,C}$ is also independent of $T$.  The dotted line shows the Gaussian approximation Eq.~\eqref{eq:prefactor_Gauss}.  The dotted vertical line indicates $q = 1/R_\mr{g} \simeq 0.69$ ($R_\mr{g} \simeq 1.45$).} 
\efig
Figure~\ref{fig:S_sC_over_w_squared_T046+T1.0+Gauss} compares $w_\mr{m,C}(q)^2/w(q)^2$ with the Gaussian approximation [see \cite{Comment_on_monomer_distr_in_CM_frame} and Eq.~(\ref{eq:Debye})]
\begin{equation}
\frac{w_\mr{m,C}^\mr{G}(q)}{w_\mr{D}(q)} =
\frac{\sqrt{\pi}\,(qR_\mr{g})^3\, \mr{e}^{-q^2R_\mr{g}^2/12} \, \text{erf}\,\big (qR_\mr{g}/2\big)}{2(\mr{e}^{-q^2R_\mr{g}^2} + q^2R_\mr{g}^2 -1)} \;.
\label{eq:prefactor_Gauss}
\end{equation}
Quantitatively, Eq.~\eqref{eq:prefactor_Gauss} is not very accurate, presumably because our chains are too short.  (The results for semi-dilute solutions of long chains obtained in Ref.~\cite{KrakoviackEtAl:EPL2002} appear to agree better with Eq.~\eqref{eq:prefactor_Gauss}.)  Qualitatively however, the Gaussian approximation reproduces the simulation results.   It starts at 1, has a maximum around $q \approx 1.5$, and vanishes for $q \gtrsim 5$.  Thus, the factor $w_\mr{m,C}(q)^2/w(q)^2$ eliminates the contributions coming from the local liquid-like structure of the melt (i.e., from $h(q)$, see Fig.~\ref{fig:S_dist_q_allT}) and, along with that, a possible dependence of $S_\mr{C}(q)$ on $T$ in our model.

\subsubsection{Correlation between the monomers and the CM}
\label{subsubsec:statics_monCM}

Figure~\ref{fig:Sa_C_s+d+all+gauss_T0.47} shows the site-resolved monomer-CM structure factors.  We see that the intra-chain contribution $w_{a,\mr{C}}(q)$ decays more slowly for the middle monomer ($a=5$) than for the end monomer ($a=1$).  This observation may be rationalized by the Gaussian approximation \cite{SchaeferKrueger_JPhys1988,Yamakawa:1971}
\begin{equation}
w_{a,\mr{C}}^\mr{G}(q) = \exp\left\{-\frac{q^2 R_\mr{g}^2}{3} 
\left[1 - 3 \frac aN  + 3 \left(\frac aN \right)^2 \right]
\right\} \;.
\label{eq:S_aC_s_gauss_in_text}
\end{equation}
Equation~\eqref{eq:S_aC_s_gauss_in_text} is symmetric under $a \leftrightarrow N-a$.  That is, chain ends are indistinguishable.  The argument of the exponential is a parabola with a minimum at $N/2$.  Thus, in qualitative agreement with the simulation data, $w_{a,\mr{C}}^\mr{G}(q)$ decreases more slowly with increasing $q$ for the middle monomer than for the end monomers.  In real space, this implies that the middle monomer is on average closer to the CM than the chain end \cite{Yamakawa:1971}.

\bfig 
\rsfig{Sa_C_s+d+all+gauss_T0.47.eps}
\caption[]{
\label{fig:Sa_C_s+d+all+gauss_T0.47}
Site-resolved structure factors resulting from monomer-CM correlations [Eqs.~(\ref{eq:def_SaC}--\ref{eq:defSqaCd})]: $w_{a,\mr{C}}(q)$ (intra-chain part), $\rho h_{a,\mr{C}}(q)$ (inter-chain part), and $S_{a,\mr{C}}(q)$ (all chains).  These structure factors are (almost) independent of $T$.  The data shown were obtained at $T=0.47$.  Circles indicate the Gaussian approximation [Eq.~\eqref{eq:S_aC_s_gauss_in_text}] for $w_{a,\mr{C}}(q)$ at $a=1$ (chain end) and $a=5$ (middle monomer).  The thickness of the lines and the symbols increases from $a=1$ to $a=5$.}
\efig

Figure~\ref{fig:Sa_C_s+d+all+gauss_T0.47} also shows the pair-correlation function $h_{a,\mr{C}}(q)$ and the sum of intra- and inter-chain contribution, $S_{a,\mr{C}}(q)$.  $S_{a,\mr{C}}(q)$ is related to the probability of finding a monomer at a certain distance from the CM of some chain, while $h_{a,\mr{C}}(q)$ measures this probability if the CM belongs to a different chain than that of the monomer $a$.  Qualitatively, $\rho h_{a,\mr{C}}(q)$ appears to be the mirror-image of $w_{a,\mr{C}}(q)$ with respect to the $q$-axis so that $S_{a,\mr{C}}(q)$ is small.  This agrees with the naive expectation that there is little correlation between the positions of the monomers and the CM's.  However, $S_{a,\mr{C}}(q)$ is not completely structureless.  It exhibits a maximum for the middle monomer ($a=5$), but a minimum for the end monomer ($a=1$).  Quite surprisingly, we find a positive correlation of the middle segments and the CM's.  On average, the probability of finding a middle (an end) monomer around the CM of a chain is increased (decreased) relative to random packing.  Intra-molecular correlations are thus canceled by inter-molecular ones only at large distances (small $q$).  At intermediate distances the intra-chain density distribution is either too little or too strongly compensated by the surrounding polymers.  By averaging over all monomers along the backbone of the chain this site-dependence of $S_{a,\mr{C}}(q)$ is suppressed to a large extent.  This means that a PRISM-like theory using the monomer averaged $S_\mr{m,C}(q)$ (see \figref{fig:S_mC_T0.47}) only, could underestimate the monomer-CM coupling.

\bfig 
\rsfig{h_aC+PRISM_T0.47.eps}
\caption[]{\label{fig:haC+PRISM}
Comparison of the monomer-CM inter-molecular pair-correlation function $h_{a,\mr{C}}(q)$ determined from the simulation data at $T=0.47$ for $a=1$ and $a=5$ (circles) and from the PRISM theory (Eq.~\eqref{eq:PRISM-haC}, solid lines).  The data for $h_{1,\mr{C}}(q)$ are shifted vertically for clarity.}
\efig

However, this does not imply that the PRISM theory cannot be applied to explain the site-dependence of $h_{a,\mr{C}}(q)$.  Equation~\eqref{eq:krak_approx2_in_text} results from the assumption that the CM may be treated as an additional, non-interacting site in the PRISM approach.  That is, the monomer-CM- and the CM-CM direct correlation functions are supposed to vanish; only $c(q)$ is kept.  Using this assumption and $S_\mr{m,C}/S=w_\mr{m,C}/w$ (see Appendix~\ref{app:krakoviack}) we find from Eq.~\eqref{eq:RISM}
\begin{equation}
h_{a,\mr{C}}(q) = \frac{w_\mr{m,C}(q)}{w(q)^2} 
\bigg [\sum_{x=1}^N w_{ax}(q) \bigg ] \, h(q)\;.
\label{eq:PRISM-haC}
\end{equation}
Figure~\ref{fig:haC+PRISM} illustrates that Eq.~\eqref{eq:PRISM-haC} is in good agreement with the simulation data.  This allows two conclusions:  First, \figref{fig:haC+PRISM} emphasizes again that the structural properties of our model may be understood in terms of the site-independent inter-chain direct correlation function and site-dependent intra-chain structure factors.  Second, the finding that $h_{a,\mr{C}}(q)$ does not depend on temperature for our model (Fig.~\ref{fig:Sa_C_s+d+all+gauss_T0.47}) may be explained by the same argument put forward for $S_\mr{C}(q)$.  It is related to the intra-chain contribution $ w_\mr{m,C}/w(q)$ which is independent of $T$ and suppresses the temperature dependence of $h(q)$ for wave vectors around $q_\mr{max}$ (see Fig.~\ref{fig:S_sC_over_w_squared_T046+T1.0+Gauss}).

\section{Results on Three-Particle Correlation Functions}
\label{sec:results_s3}

Recently, triple correlations in simple and network glass-forming liquids have been investigated \cite{SciortinoKob2001}.  This study shows that, while the convolution approximation (Eq.~\eqref{eq:S_3_and+c_3} with $c_3=0$) is very good for simple liquids, it fails to provide an accurate description of the cage structure in silica.  As silica is a network-forming liquid, nearest-neighbor bonds make an important contribution to the local structure in the liquid.  This is similar to the chain connectivity in a polymer melt.  So, we investigate the importance of triple correlations for our model by comparing the three-monomer structure factor with its convolution approximation for selected subsets of $(q,k,p)$.

\bfig 
\rsfig{S_3_qqq+convolution_approx.eps}
\caption[]{\label{fig:S_3_qqq+convolution_approx}
Comparison of the three-monomer structure factor (thin lines) for the melt, $S_3(q,q,q)$, and for the polymers, $w_3(q,q,q)$, with the respective convolution approximations (thick lines), $S(q)^3$ and $w(q)^3$ at $T=0.47$.  The simulation results for the triple correlations are not smoothed.  The lower statistical accuracy of $S_3(q,q,q)$ compared to $S(q)^3$ is clearly visible, especially at large $q$.} 
\efig

Figure~\ref{fig:S_3_qqq+convolution_approx} presents the three-monomer structure factor of the melt, $S_3(q,q,q)$, and of the chains, $w_3(q,q,q)$, for the choice that the three vectors $\vec{q}$, $\vec{k}$, and $\vec{p}$ form an equilateral triangle characterized by the length of its side $q$.  We find that the convolution approximation provides a very good description for the triple correlation of the polymers, the amplitude of the oscillations being slightly underestimated, however.  For $q \lesssim 20$, $S_3(q,q,q)$ is equally well represented by the convolution approximation, expect for a sharp dip at $q \approx 6.3$, revealing some anti-correlation at this $q$-value.  For $q \gtrsim 20$ the interpretation of the data is difficult due to the high noise level.  In this region, we find that $S_3(q,q,q)$ is systematically larger than $S(q)^3$ and even stays above unity, the theoretical large $q$-limit of both quantities.  This difference must be attributed to insufficient statistics \cite{Comment_on_S3}.  Despite this proviso, the convolution approximation represents a fairly good description of the three-monomer correlations for the choice of wave vectors ($q,q,q$).   

\bfig 
\rsfig{S_p_3_q_cos_phi_isoceles+conv_approx.eps}
\rsfig{S_3_q_cos_phi_isoceles+conv_approx.eps}
\caption[]{
\label{fig:S_3_q_cos_phi_isoceles+conv_approx}
$w_3(q,q,p=q\sqrt{2(1-\cos\varphi)})$ (top) and $S_3(q,q,p=q\sqrt{2(1-\cos\varphi)})$ (bottom) versus $\cos \varphi$ for some selected $q$-values at $T=0.47$.  The simulation data for $w_3$ and $S_3$ are represented by thin lines, the convolution approximation [Eq.~\eqref{eq:S_3_and+c_3} with $c_3=0$] by thick lines.  Note that the data for $S_3(q)$ at $q=7.1$ are rescaled by a factor of 0.1.} 
\efig

In order to investigate the angular dependence of the triple correlations we follow a suggestion made in Ref.~\cite{BarratHansenPastore1987}.  We determine $S_3$ and $w_3$ for the triple of moduli $(q,k=q,p=q\sqrt{2(1-\cos\varphi)})$, i.e., for isosceles triangles with two sides of length $q$ enclosing an angle $\varphi$ [Eq.~\eqref{phi_qk}].  

Figure~\ref{fig:S_3_q_cos_phi_isoceles+conv_approx} shows the simulation results and the convolution approximation as a function of $\cos\varphi$ for various $q$ corresponding to maxima and minima positions of the three point structure factors (cf.\ Fig~\ref{fig:S_3_qqq+convolution_approx}).  As found before, the agreement between $w_3$ and the convolution approximation is very good, except at $q=24.9$ where the approximation yields oscillations that are absent in $w_3$.   Similarly for most $q$-values, $S_3$ and its convolution approximation are fairly close to one another.  Barring $q=24.9$, for which the quality of the comparison is hard to judge due to the noise in $S_3$, noticeable deviations are obtained for wave vectors close to $q_\mr{max}$.  This might suggest that the cage structure in the cold melt imposes triple correlations which are different than those predicted by the convolution approximation.  To test this conjecture the statistics of the data should be improved considerably, which is currently hard to achieve \cite{Comment_on_S3}.

\section{Summary and Conclusions}
\label{sec:mct_conclusions}

We explored static properties of a supercooled, non-entangled polymer melt consisting of flexible chains.  The temperatures studied range from the high-$T$, normal liquid state of the melt to the supercooled state close to, but above the critical temperature of MCT ($\Tc > \Tg$).  Our analysis utilizes various structure factors characterizing spatial correlations, on different length scales, between the monomers, between the monomers and the CM's, and between the CM's.  The main findings of our work may be summarized as follows:

Due to the flexiblity of the chains in our model (e.g., no bond-angle or torsional potential), their conformational properties remain essentially unchanged on cooling.  Thus, all intra-chain structure factors, $w(q)$, $w_{ab}(q)$ and $w_{a,\mr{C}}(q)$, only depend very weakly on $T$ (Figs.~\ref{fig:w_q+approx_all_T},\ref{fig:w_ab_T0.47_1.1-1.10+inner},\ref{fig:Sa_C_s+d+all+gauss_T0.47}).

However, not only the intra-chain, but also the inter-chain correlation functions involving the CM, $S_\mr{C}(q)$ and $h_{a,\mr{C}}(q)$, are (almost) independent of $T$ (Figs.~\ref{fig:S_chain_cm_all_T+krakoviack_approx},\ref{fig:Sa_C_s+d+all+gauss_T0.47}).  We explain this finding by PRISM theory which relates $S_\mr{C}(q)$ and $h_{a,\mr{C}}(q)$ to the intra-chain structure by a term containing $w_\mr{m,C}/w(q)$.  This ratio is (almost) independent of $T$ (\figref{fig:S_sC_over_w_squared_T046+T1.0+Gauss}) and suppresses the temperature dependence of the inter-chain correlations on the local scale of the nearest-neighbor shells.

The CM-CM structure factor $S_\mr{C}(q)$ is fairly featureless (\figref{fig:S_chain_cm_all_T+krakoviack_approx}).  Outside the small-$q$ regime reflecting the low compressibility of the melt, $S_\mr{C}(q)$ quickly approaches the ideal gas behavior, $S_\mr{C}(q)=1$.  This shows that the chains are soft, interpenetrating objects whose spatial arrangement does not change on cooling.  For our model, the slowing down of the CM motion, observed in previous works \cite{BennemannPaulBaschnagel1999_Rouse}, is thus not related to a tighter packing of the polymers at low $T$.

Correlations between the monomers are reflected by the collective structure factor $S(q)$.  $S(q)$ exhibits liquid-like oscillations which become more pronounced with decreasing temperature (\figref{fig:S_q_melt_allT}).  This signature of a tighter packing in the nearest-neighbor shells results from the inter-chain contribution $h(q)$ (\figref{fig:S_dist_q_allT}).  Thus, the sluggish dynamics of our model is driven by the nearest neighbors that are not directly bonded to each other.  In this respect, our polymer melt corresponds to a simple glass-forming liquid.  

Another result supports this correspondence.  The convolution approximation which factorizes the three-particle structure factors $S_3(q,k,p)$ in the product $S(q)S(k)S(p)$ is generally invoked in the mode-coupling theory for the glass transition.  Several studies show that this approximation is well justified for simple glass formers \cite{SciortinoKob2001,BarratGoetzeLatz1989}, but not for structurally more complicated ones, such as {\em ortho}-terphenyl \cite{RinaldiSciortinoTartaglia2001} or silica \cite{SciortinoKob2001}.  In our model we find, analogously to simple liquids, that the convolution approximation works quite well, except for $S_3(q,k,p)$ at intermediate $q$ (close to $q_\mr{max}$) and for certain angles between the wave vectors (\figref{fig:S_3_q_cos_phi_isoceles+conv_approx}).  To what extent these deviations could be important in a mode-coupling calculation is hard to estimate quantitatively (due the lack of sufficient statistics \cite{Comment_on_S3}).

The analysis of the monomer-resolved structure factors shows that the intra-chain contribution $w_{ab}(q)$ depends, to a very good approximation, only on the distance $|a-b|$ between the monomers $(a,b)$ along the backbone.  Thus, chain end effects are not very important for the intra-molecular structure of our model.  On the other hand, the inter-chain structure factor $h_{ab}(q)$ depends expli\-citly on the monomer pair $(a,b)$.  The site-dependence of $h_{ab}(q)$ and of $h_{a,\mr{C}}(q)$ may be explained by PRISM theory which assumes that the direct correlation function is independent of the monomer index (Figs.~\ref{fig:h-site},\ref{fig:haC+PRISM}).  By calculating the site-site and site-averaged direct correlation functions we test this assumption and verify that it represents a good approximation (\figref{fig:c-site}).  This shows that the structural properties of our model, even subtle monomer-monomer- and monomer-CM correlations, may be calculated from the site-averaged inter-chain direct correlation function and the site-dependent intra-chain structure factors, both of which are determined in the simulation.  A similar agreement between PRISM theory and computer simulations of coarse-grained polymer models was also found in other studies \cite{KrakoviackEtAl:EPL2002,YethirajEtal:JCP1990,CurroSchweizerGrestKremer1989,SchweizerHonnellCurro:JCP1992}.  In a forthcoming article, we will exploit the results of the present work to propose and test a mode-coupling theory for the dynamics of our supercooled polymer melt.

 
\begin{acknowledgments}
We are indebted to Dr.\ H. Meyer for many helpful discussions and to Prof.\ K. Binder for a critical reading of the manuscript.  M. F. thanks the members of the theory group and the staff of the Institut Charles Sadron (ICS) for the warm hospitality he enjoyed while visiting the ICS.  This work was made possible by generous grants of computer time at the com\-put\-er center of the Johannes Gutenberg-Universit\"at Mainz and at the IDRIS in Orsay.  Financial support by the DAAD (grant No.\ D/00/07994), the BMBF (grant No.\ D.I.P.\ 352-101 and grant No.\ 03N 6500), the ESF SUPERNET Prog\-ramme, the DFG (grant No.\ Fu 309/3), the DFG and the MENRT (IGC ``Soft Matter''), and the ``Institut Universitaire de France'' is gratefully acknowledged.
\end{acknowledgments}
 

\appendix
\section{PRISM Approximation for the Structure Factor of the Chain's Center of Mass}
\label{app:krakoviack}

The recent work by Krakoviack {\em et al.}\ \cite{KrakoviackEtAl:EPL2002} uses PRISM theory to calculate the CM structure factor $S_\mr{C}(q)$ from monomer correlation functions.  This approach was found to compare well with simulation data of (long) chains in semi-dilute solution.  For polymer melts Krakoviack {\em et al.}\ did not test the theory and even mention the caveat that the correlation between the CM and the monomers could be different.  Here, we suggest that their theory also yields a reasonable approximation of $S_\mathrm{C}(q)$ for polymer melts.  In the following, we sketch the main ideas of the approach of Ref.~\cite{KrakoviackEtAl:EPL2002} and test the basic assumptions against our simulation results.

The starting point of Ref.~\cite{KrakoviackEtAl:EPL2002} consists in introducing the center of mass as an additional, non-interacting site.  This step generalizes the functions $h$, $c$, $w$, and $S$ [Eqs.~(\ref{eq:Sq}--\ref{eq:def_hq},\ref{eq:PRISM-S1})] to $2\times 2$ matrices
\begin{align}
  \matr{h} & = \left(
  \begin{array}{cc}
    N h_\mr{m,m} & \sqrt{N} h_\mr{m,CM}\\
    \sqrt{N} h_\mr{m,CM} & h_\mr{CM,CM}
  \end{array}
\right)\;, \label{eq:def_mat_h}\\
\matr{c} & = \left(
  \begin{array}{cc}
    N c_\mr{m,m} & \sqrt{N} c_\mr{m,CM}\\
    \sqrt{N} c_\mr{m,CM} & c_\mr{CM,CM}
  \end{array}
\right)\;, \label{eq:def_mat_c}\\
\matr{w} & = \left(
  \begin{array}{cc}
    w_\mr{m,m}& \frac{1}{\sqrt{N}}w_\mr{m,CM}\\
    \frac{1}{\sqrt{N}}w_\mr{m,CM} & w_\mr{CM,CM}
  \end{array}
\right)\;,\\
\matr{S} & =  \left(
  \begin{array}{cc}
    S_\mr{m,m}& \frac{1}{\sqrt{N}}S_\mr{m,CM}\\
    \frac{1}{\sqrt{N}}S_\mr{m,CM} & S_\mr{CM,CM}
  \end{array}
\right)\;.
\end{align}
Here, the indices ``m,m'', ``m,CM'', and ``CM,CM'' denote the monomer-monomer, monomer-CM, and CM-CM correlations, respectively.  By definition, all matrices are symmetric.  With the notation of Sect.~\ref{subsec:dens_corr} we have the following identities
\begin{align}
h_\mr{m,m} & = h\,, \;\; h_\mr{m,CM}=h_\mr{m,C}\,, \;\;
h_\mr{CM,CM} = h_\mr{C} \;, \nonumber \\
w_\mr{m,m} & = w\,, \;\; w_\mr{m,CM}=w_\mr{m,C}\,, \;\;
w_\mr{CM,CM}=1\;, \label{eq:hwS_identities} \\
S_\mr{m,m} & = S\,, \;\; S_\mr{m,CM}=S_\mr{m,C}\,, \;\; 
S_\mr{CM,CM}=S_\mr{C} \;. \nonumber 
\end{align}

The matrices of correlation functions are related to each other by Eqs.~(\ref{eq:def_cab},\ref{eq:PRISM-sh}), i.e.,
\begin{align}
\rho \matr{c}(q) &= \matr{w}^{-1}(q)-\matr{S}^{-1}(q) \;, \label{eq:prism_c} \\
\matr{h}(q) &= \matr{w}(q)\matr{c}(q)\left[\matr{w}(q) + 
\rho \matr{h}(q)\right]\;. \label{eq:prism_h}
\end{align}
Using Eq.~\eqref{eq:hwS_identities} we find from Eq.~\eqref{eq:prism_c}
\begin{align}
  \rho_\mr{m} c_\mr{m,m} & =\frac{1}{w - w_\mr{m,C}^2/N} - \frac{1}{S - S_\mr{m,C}^2/(N S_\mr{C})}\;,\nonumber\\
  \rho_\mr{m} c_\mr{m,CM} &= \frac{-w_\mr{m,C}}{w - w_\mr{m,C}^2/N} + \frac{S_\mr{m,C}}{S_\mr{C} S - S_\mr{m,C}^2/N}\;,\label{eq:c_XY} \\
  \rho c_\mr{CM,CM} &= \frac{1}{1 - w_\mr{m,C}^2/(N w)} - \frac{1}{S_\mr{C} - S_\mr{m,C}^2/(N S)} \;. \nonumber
\end{align}
Note that $\rho_\mr{m} = N \rho$ [Eq.\eqref{eq:defdensities}].  In Ref.~\cite{KrakoviackEtAl:EPL2002} it is assumed that
\begin{equation}
\label{eq:assum_cs}
c_\mr{CM,CM} \equiv 0\; \quad c_\mr{CM,m} = c_\mr{m,CM} \equiv 0\;,
\end{equation}
and only $c_\mr{m,m} \neq 0$ is retained.  This implies that the centers of mass interact neither with each other nor with the monomers.

\bfig 
\rsfig{all_dcf_T0.47_new.eps}
\caption[]{
\label{fig:all_dcf_T0.47}
Direct correlation functions versus $q$ at $T=0.47$.  The correlation functions $c_\mr{m,m}(q)$, $c_\mr{m,CM}(q)$, and $c_\mr{CM,CM}(q)$ are calculated from Eq.~(\ref{eq:c_XY}).  The results of this calculation are numerically reliable for $q \gtrsim 1$ \cite{Comment_on_matrix_OZ_equation}.  Additionally, the figure shows the monomer-monomer direct correlation $c(q)$ given by Eq.~\eqref{eq:PRISM-S1} and the CM-CM direct correlation function $c_\mathrm{C}(q)$ defined from $S_\mathrm{C}(q)$ by $\rho c_\mr{C}(q) = 1 - 1/S_\mr{C}(q)$.  The monomer-monomer direct correlation functions $c_\mr{m,m}(q)$ and $c(q)$ are indistinguishable from one another (for $q \gtrsim 1$ \cite{Comment_on_matrix_OZ_equation}).  For $q\rightarrow 0$, $\rho_\mr{m}c(q)$ and $\rho c_\mr{C}(q)$ tend to the limits [see Eqs.~(\ref{eq:PRISM-S1},\ref{eq:SC_limit_q_to_0})]: $N\rho_\mr{m}c(q\rightarrow 0)= 1-N/k_\mr{B}T\rho_\mr{m} \kappa_T = \rho c_\mr{C}(q\rightarrow 0)$ ($\approx -840$ for $T=0.47$).} 
\efig

To test the validity of this assumption we calculated the right-hand side of Eq.~\eqref{eq:c_XY} from the simulation data and checked whether the direct correlation functions involving the CM are small compared to $c_\mr{m,m}$.  Figure~\ref{fig:all_dcf_T0.47} shows that the monomer-monomer direct correlation function is indeed (much) larger than the other components of $\matr{c}$, at least for $q \gtrsim 1$ \cite{Comment_on_matrix_OZ_equation}.  Thus, Eq.~\eqref{eq:assum_cs} appears to be a reasonable approximation for our model in the $q$-range, where $S_\mr{C}(q)$ substantially deviates from its small-$q$ limit, Eq.~(\ref{eq:SC_limit_q_to_0}) (for a more precise test see \figref{fig:S_mC_T0.47}).

Inserting Eq.~\eqref{eq:assum_cs} into the Ornstein-Zernike equation for $\matr{h}$ [Eq.~\eqref{eq:prism_h}] and using Eqs.~(\ref{eq:hwS_identities},\ref{eq:Sq}) we recover Eq.~\eqref{eq:PRISM-S1} for $c_\mr{m,m}(q)$ ($= c(q)$).  Furthermore, if we insert Eq.~\eqref{eq:assum_cs} into Eq.~\eqref{eq:c_XY}, we obtain the relation $S_\mr{m,C}=w_\mr{m,C}S/w$ (for a test see Fig.~\ref{fig:S_mC_T0.47}) from the second and the third line of Eq.~\eqref{eq:c_XY}.  Using this relation and Eq.~\eqref{eq:PRISM-S1} in the first line of Eq.~\eqref{eq:c_XY} we find 
\begin{equation}
\label{eq:krak_approx2}
\begin{split}
S_\mr{C}(q) & = 1 + \frac{1}{N}\,\frac{w_\mr{m,C}(q)^2}{w(q)^2} \, \big [S(q) - w(q) \big ] \\
 & = 1 + \frac{1}{N}\,\frac{w_\mr{m,C}(q)^2}{w(q)^2} \, \rho_\mr{m} h(q) \;,
\end{split}
\end{equation}
i.e., Eq.~\eqref{eq:krak_approx2_in_text}.  Equation~\eqref{eq:krak_approx2} is the central result obtained in Ref.~\cite{KrakoviackEtAl:EPL2002} (see Eq.~(16) of  Ref.~\cite{KrakoviackEtAl:EPL2002}).

\bfig
\vspace{4mm}
\rsfig{S_mC+PRISM_T0.47.eps}
\vspace{-4mm}
\caption[]{
\label{fig:S_mC_T0.47}
Average monomer-CM static structure factor $S_\mr{m,C}(q)$ [Eq.~\eqref{eq:defSqaC}] versus $q$.  The solid line represents the simulation results at $T=0.47$.  The dashed line is the PRISM prediction $S_\mr{m,C}^\text{PRISM}(q)=w_\mr{m,C}(q)S(q)/w(q)$, where the right-hand side was calculated from the simulation.  Although the simulated $S_\mr{m,C}(q)$ is fairly noisy, the comparison still suggests that there are systematic deviations between $S_\mr{m,C}(q)$ and $S_\mr{m,C}^\text{PRISM}(q)$ for $q \lesssim 6.5$.  These deviations come from the assumption of Eq.~\eqref{eq:assum_cs}: the direct correlation functions involving the CM become truly 0 only for $q \gtrsim 6.5$.  The deviations might be responsible for the differences found between the simulated $S_\mr{C}(q)$ and \eqref{eq:krak_approx2} at $q \approx 3$ [see \figref{fig:S_chain_cm_all_T+krakoviack_approx}; we could not test this hypothesis due to insufficient statistics of $S_\mr{m,C}(q)$].  However, they do not appear to hamper the good agreement between the simulation results and the PRISM predictions in Figs.~\ref{fig:S_chain_cm_all_T+krakoviack_approx} and \ref{fig:haC+PRISM}.} 
\efig


\bibliography{references_Polymer,references_Glass_MCT_and_LiquidTheory,references_DynamicHeterogeneity}

\begin{thebibliography}{61}
\bibitem[{Com({\natexlab{c}})}]{Comment_on_PRISM}
\bibinfo{note}{The PRISM results are obtained in the following way: First,
  $c(q)$ is calculated by Eq.~(\ref{eq:PRISM-S1}) using $S(q)$ and $w(q)$ from
  the simulation. The resulting $c(q)$ and the matrix $w_{ab}(q)$ are then
  inserted into Eq.~(\ref{eq:PRISM-sh}), assuming $c_{ab}(q) = c(q)$, to find
  $h_{ab}^{\rm PRISM}(q)$. Equivalently, Eq.~\eqref{eq:PRISM-h} can be used
  with the simulation data for $h(q)$, $w(q)$, and $w_{ab}(q)$.}
\bibitem[{Com({\natexlab{e}})}]{Comment_on_matrix_OZ_equation}
\bibinfo{note}{Equation~(\ref{eq:prism_h}) requires that the inverse
  ma\-tri\-ces, $\matr{w}^{-1}(q)$ and $\matr{S}^{-1}(q)$, exist. To this end,
  we must have $\det \matr{w}(q) = w(q) - [w_\mr{m,C}(q)]^2/N \neq 0$ and
  $\det\matr{S}(q) = S(q)S_\mr{C}(q) - [S_\mr{m,C}(q)]^2/N \neq 0$. Because
  $w(q \rightarrow 0)=w_\mr{m,C}(q\rightarrow 0)=N$ and $\lim_{q\rightarrow
  0}S(q)S_\mr{C}(q)=(k_\mr{B}T\rho_\mr{m}\kappa_T)^2/N=\lim_{q\rightarrow
  0}[S_\mr{m,C}(q)]^2/N$, the determinants vanish in the limit $q\rightarrow
  0$. In our model, the structure factors begin to approach their respective
  small-$q$ behavior for $q < 1$. Thus, we expect the results for the direct
  correlation functions obtained from Eq.~(\ref{eq:prism_c}) to become
  numerically imprecise for $q < 1$.}
\end{thebibliography}

\end{document}